\documentstyle[prb,twocolumn,aps]{revtex}

\begin{document}
\tolerance 50000

\draft

\title{
Snakes and Ladders}
\author{M. A. Mart{\'\i}n-Delgado$^{1}$, R. Shankar$^{2}$ and
 G. Sierra$^{3}$
}
\address{$^{1}$Departamento de F{\'\i}sica
Te{\'o}rica I. Universidad Complutense.  28040-Madrid, Spain \\
$^{2}$Sloane Physics Laboratories, Yale University, New Haven CT  06520 \\
$^{3}$ Instituto de Matem{\'a}ticas y F{\'\i}sica Fundamental.
C.S.I.C. Serrano 123, 28006-Madrid, Spain
}

\twocolumn[

\date{April 96}
\maketitle
\widetext

\vspace*{-1.0truecm}

\begin{abstract}
\begin{center}
\parbox{14cm}{
We map spin ladders with $n_l$ legs 
and  couplings $J'$ across all 
 rungs and  $J(1 \pm \gamma)$ along the legs, 
 staggered  in both directions,  
 to a sigma model.  Setting its $\theta = (2m+1)\pi$
(where it is known to be gapless), we  locate   the critical curves in the $\gamma$   
versus ${J'\over J}$ plane  at each $n_l$, and spin $S$.  
The phase diagram is rich and has  some surprises: when 
   two gapped  chains are suitably  coupled,
 the combination becomes gapless. With $n_l, \gamma $ and $J'/J$  
to control, the prospects for experimentally observing any one of 
these equivalent transitions  seem 
bright.    
We discuss the order parameters and the behavior of holes in the RVB description.
}
\end{center}
\end{abstract}

\pacs{
\hspace{1.9cm}
PACS numbers: 75.10.Jm, 75.50.Ee, 75.30.Ds}

]

\narrowtext
With  this paper we 
contribute to the explosive growth in the  
theoretical and experimental studies  of antiferromagnetic spin chains  
and ladders. 
Let begin with   Haldane's \cite{Hal} mapping of the spin-$S$ Heisenberg  
chain with hamiltonian 

\begin{equation}
H = J \sum_n {\bf S}(n) \cdot {\bf S}(n+1) \label{1}
\end{equation}
via spin coherent states 
\noindent 
on to the nonlinear sigma model with euclidean action 
\begin{equation}
S = \int dx d\tau \left[ - {1 \over 2 g} (\nabla {\bf  \Phi} )^2 + 
i {\theta \over 
4\pi }  {\bf  \Phi} \cdot \partial_x  {\bf  \Phi} \times \partial_{\tau} {\bf  \Phi} \right]. 
\end{equation}
Here $\Phi$ is a three component unit vector, and   
 $\theta$,  which multiplies $i$ 
times the integer valued winding number $W$,  is

\begin{equation}
\theta = 2 \pi S. \label{2}
\end{equation} 
There is  a  hamiltonian derivation of this 
result due to Affleck\cite{AffLH} which we shall allude to later.

Since $\theta$ enters the path integral via $e^{i\theta W}$,   
it matters only  mod $2\pi$ and   
when $\theta =0,\ \pi$, the path integral 
is invariant under $x \to -x$ (parity) under which $W \to -W$.   
\noindent
One now argues  that integer spin chains have a gap  
and half-integer spin chains do not. First,   all  
integer (half-integer) spin chains essentially have $\theta =0$ ($\pi $).
  The  sigma model with $\theta =0$   
is known to have exponentially decaying correlations \cite{HKS}.   
As for  $\theta =\pi $, since  the spin-${1 \over  
2}$  Bethe  chain is gapless,  so must be  the $\theta =\pi$ sigma model  
, provided the mapping to  the sigma model
 (most reliable for large $S$) is valid down to spin ${1 \over 2}$.
 Since so much in this paper hinges 
on the masslessness of the $\theta = \pi$ model,  we
point out   that 
 Shankar  
and Read \cite{Shan} have shown independently that the sigma model at $\theta = 
\pi$ is massless by considering the 
$\tau$-continuum hamiltonian of the lattice regulated model.

It is also accepted that nonstaggered ladders with $n_l$ legs are  
gapless only if $n_lS$ is half-integer\cite{Dag,Bar,White,Gopalan,DaRi}.
 This is most transparent  when  
the interleg coupling  $J'$ is much larger than the intraleg coupling  
$J$, for  we can first 
 solve the problem of $n_l$ spins along a rung, 
take  the lowest energy multiplet in  
each rung, and then couple them with $J$, thereby getting a  
single chain, about which everything is known.
 One then verifies that nothing changes as $J'$ is lowered. Equivalently  
\cite{Kh,sierra} one can show  that the topological terms 
for the chains are additive, giving $\theta  = 2 \pi n_l S$. 
Different modifications of spin ladders can also
be considered by adding next-to-nearest neighbor  couplings \cite{White2,breh}.
Coupled spin chains have been studied  by
combining mean field theory techniques with exact results for
one chain \cite{schulz}.

The spin systems  to be  
considered here have a very 
important feature: {\em they  have staggered weights}. Let us then begin with  
a single chain for which

\begin{equation}
J(n) = J (1 + (-1)^n\gamma ) \label{3}
\end{equation}
is the coupling between sites $n$ and $n+1$. 
 Notice that  $\gamma \to -\gamma$ amounts to sublattice exchange
 $n \to n+1$  and that the restriction 
 $|\gamma |<1$, keeps the interaction antiferromagnetic.

 Affleck and Haldane
  \cite{AH,AffLH} 
showed that in this case  
\begin{equation}
\theta = 2 \pi S ( 1 + \gamma ) \label{4}
\end{equation}
so that  when $\gamma$ is varied from $-1$ to $+1$, $\theta$  passes  
an odd multiple of $\pi $  i.e., the system is critical,  exactly $2S$ times.
It is instructive to interpret  these transitions in the 
valence bond terminology of Affleck et al. (AKLT) \cite{AKLT}, wherein each
  spin-S is  
viewed as a symmetrized product of $2S$ spinors.  As $\gamma$ is  
raised from $-1$, the chain goes from 
being  fully dimerized with all the valence   
bonds (spinor contractions) on  odd-$n$  links, to being dimerized  with all  
valence bonds on the even-$n$  links. 
As each spinor index switches loyalty, 
it necessarily reaches a point when it can equally well go either  
way, producing a nonstaggered i.e., 
a gapless spin-${1 \over 2}$  chain. (It is important
 to realize that the {\em effective}  interaction of these spin-${1\over 2}$
 degrees of freedom can be nonstaggered even though the
original  Heisenberg interaction is.)

Consider  staggered chains shown in Fig. \ref{f1} 
  with    
 horizontal  couplings on the $a^{\text{th}}$-leg 
($a=1, \ldots, n_l$) obeying  

\begin{equation}
J_a (n) = J (1 + (-1)^{n+a+1} \gamma)  \label{5}
\end{equation}

\noindent 
i.e.,  staggered in both directions. We now show that  such systems  
have a rich  phase structure in the $\gamma$ versus  
$ J'/J$ plane at each $n_l$ and $S$.

Recall   Affleck's derivation of 
 the sigma model hamiltonian from the spin chain  
by pairing spins, forming their difference and sum,  and turning these into the 
sigma model field and its conjugate momentum respectively, 
in the limit of large $S$. This method was generalized by  
Sierra \cite{sierra} to uniform ladders. The main difference was the   
 $n_l$-fold increase in the number of degrees of freedom due  
to the transverse label $a=1, \ldots n_l$ for the legs. A low energy  
analysis indicated that  only one of these modes remained   
low in energy and defined the effective sigma model, 
while the rest had  a gap of order  
$J'$. This effective model had $\theta = 2\pi n_lS$, 
(independent of couplings), yielding the  
previously quoted result  for  
nonstaggered ladders, 
namely that only an odd number of half-integer chains were massless. 

We extended this derivation to the staggered ladders  and found  
\begin{equation}
\theta = 2 \pi S n_l ( 1 + \gamma f_{n_l}(J'/J)) \label{6}
\end{equation}

\noindent
where 

\begin{eqnarray}
f_{n_l}(J'/J) =  {1\over n_l^2} [ \delta_{n_l,{\text{odd}}} \cr
+ 2 \; \sum _{m=1,3,\ldots,n_l -1} {1\over \sin^2 ({\pi m \over 2 n_l})} 
{1\over 1 + {J^\prime \over J} \cos ^2 ({\pi m \over 2 n_l})} ] \label{7}
\end{eqnarray}

\noindent with $\delta_{n_l,{\text{odd}}}$ equal to 1 if $n_l$ is odd and zero 
otherwise.
We   
refer the reader to Sierra\cite{sierra}  for a very similar  
derivation in the uniform case.

{\em The critical points follow from setting this $\theta$ equal to odd  
multiples of $\pi$.}

We analyze two cases: $n_l=2, S={1 \over 2},1$ which  should   convince the reader of the soundness of this  
method, and facilitate the  
discussion of  the cases with larger $S$ and $n_l$.

Consider Fig. \ref{f2}.
On the $\gamma $ 
axis, where the chains decouple, there is just one (which equals   
 $2S$) critical point corresponding to the nonstaggered  
spin-${1 \over 2}$ chain. Our theory predicts  that as we turn on  
$J'$, this becomes two critical points that move towards the walls  
$|\gamma |=1$. {\em  It also tells us that although staggering or interchain  
coupling are individually bad for criticiality, a certain combination  
can sustain  criticality.}  Can we believe this?  Let us go to $\gamma =  
-1$. Now each chain breaks up into disconnected pairs, but the  
disconnected pairs of one chain do not lie opposite  those of the other,  
but displaced by one unit. When these get coupled by $J'$, we have a  
``snake" chain that winds through the lattice. It is a spin-${1 \over  
2}$ chain with staggered weights $2J$ and $J'$. Clearly at $J'=2J$,  
it becomes critical as predicted by the theory. Thus the vertical  
$J'$ axis is seen to play  the role of an effective $\gamma$ for the snake.  
We display this by showing three snakes on the left margin of Fig. \ref{f2}, 
 with  vertical bonds  which are stronger than, equal to and weaker  than 
 the horizontal ones ( $2J$). As expected, the same thing happens on $\gamma  
= +1$, with $n     
\to n+1$.

Although it is not so easy to understand  criticality as we  
go into the rectangle, 
by continuity of $\theta$,  the critical curve must exist.  
There is however one caveat: 
the phase diagram in Fig.\ref{f2}  does not strictly  
follow from the equation for $\theta$ when   $J' \to 0$: 
 the two critical  
curves coming down from    $J'/J=2$ on $|\gamma |=1$  
 will cut the $\gamma$ axis at 
distinct points on either side of the origin 
instead of meeting there. 
But we know that the  sigma model mapping is doomed to fail
 as $J^\prime \to 0$:  we will get not one low energy field, (the putative  
sigma model field) but two,   
since the gap that separated the sigma model field from the other, of  
order $J'$, vanishes.  Fortunately, on the $\gamma$  
axis, where the chains decouple,  
we know everything: there is only one transition at $\gamma =0$  
which  the two chains undergo  simultaneously.  
Thus the final phase diagram was obtained by  
combining what we know on the $\gamma$ 
axis (about decoupled chains) with what we know off the  
$\gamma$ axis (from the sigma model)\cite{comment}.
Even off the $\gamma$ axis  the sigma model is 
only to be taken  as a guide to the
topology of the phase diagram and not for the exact
 location of the critical curves. This is because 
 the formula for $\theta$ 
is generally not exact {\em except   when $\gamma =  0$ and 
 $\theta = 2\pi n_l S$, in which case    the sigma model is
 invariant under parity and you cannot 
alter $\theta$ by a small amount 
(say of order $1/S$), without violating parity.}

Along  an arc starting at 
$\gamma = -1, \ J'=0$ and ending at 
$\gamma = 1,\  J'=0$, 
$\theta$ rises continuosly from $0$ to $4\pi$. 
The critical behavior is the same across any of these critical
 curves and the gap will behave as $t^{2/3}$, 
where $t$ is the control parameter, 
as predicted by Cross and Fisher\cite{DSF}.
(There will be logarithmic 
corrections  since the $\theta = \pi$  sigma model 
differs from the conformally invariant WZW model by a 
marginally irrelevant operator\cite{Shan}.  Chitra {\em 
et al} \cite{breh} avoid the log by adding a 
special value of nnn coupling to the spin $-{1 \over 2}$ 
chain and find an exponent  very close to $2/3$.) 

Let us examine  Fig. \ref{2} in terms of  
the RVB picture
of White, Noack and Scalapino \cite{White}   
for a nonstaggered  $n_l=2$, $S={1\over 2}$ system
which corresponds to a point vertically above the 
origin in Fig. \ref{2}, with some generic $J^\prime$.
(It might help to consult their Fig. 3 for this 
discussion.) In the absence of defects, the bonds
in each two by two square resonate between being vertical
(with coupling $J^\prime$) and horizontal 
(with coupling $J$). A defect 
 forces the bonds to be horizontal,
staggered, and nonresonating,  till we reach the next defect.
This causes a linear confining potential and restricts the
excitations to spin-$1$.
In our problem, we are free to move to the left of this
point towards negative $\gamma$.
Now the staggered horizontal bond configuration
between the defects becomes more favorable and we soon
hit the critical curve on which the staggered bonds configuration  becomes  degenerate with the 
resonant ones, 
and the defects (spinons) are liberated.
To the left of the critical curve confinement resumes,
for reasons best understood if we drop vertically from
the critical point to the $J^\prime=0$ axis.
Now we have decoupled staggered chains. The bonds
are dimerized in the preferred sublattices.
A pair of defects now forces singlets on unfavorable
bonds in the region in between. When  
$J^\prime$ is turned on,   between the defect, the bonds can resonate 
 since the defect  has 
lined them up across each other. Increasing $J^\prime$
improves resonance and we finally hit the critical 
curve. In general all our critical curves may be 
characterized as those on which the defects are unconfined.
 
 What about the order parameter for 
the different phases? Once again it is best to move up  
the $\gamma =-1$ axis, where  we see that the valence bonds go from being  
horizontal to vertical. This is just the Affelck-Haldane transfer  
of bonds on a chain, but along the length of the snake, wherein  
even/odd bonds turn into vertical/horizontal bonds.

Consider Fig. \ref{f3} for spin-$1$.
 Once again on  $\gamma = -1$ we get a  spin-$1$ snake,   
which  becomes gapless when its staggering  equals 
$\pm 1/2$ according to the sigma model (\ref{6}), \cite{comment2}. 
Once again the ratio of couplings  $J^\prime/ 2J$, determines  the
 effective staggering along the snake, not to be confused
with the original $\gamma$ for the ladder. Setting $J^\prime /2J$ equal to 
 $(1\pm   
{1 \over 2})/(1\mp {1 \over 2})$ 
we get critical values  $J'/2J = 3, 1/3$. 
 It is  
clear that we can adapt the nonlocal order parameter of den Nijs and Rommelse
\cite{den} (rendered along 
the snake) to describe the $Z_2 $ symmetries.
We do  not discuss spin-1 further since it 
resembles spin-${1 \over 2}$ in other respects.

For larger  values of spin and $n_l$,  each 
single-chain transition on the $\gamma$ axis splits into $n_l$ transtions 
 as we turn on $J'$. The critical curves bend towards the wall
 ($|\gamma |=1$) nearest to them.  
The parameter $\theta$ rises continuously from $0$ to $4\pi n_l S$ as 
we follow the arc shown in Figure \ref{2}.  There are however some differences. 
{\em First, we get honeycomb  
ladders instead of snakes for larger $n_l$. } Next, we no longer 
have an  easy way to  
see the sigma model is even qualitatively 
correct when it locates critical  lines for us.
 However, we expect the model  to  
be weakest when $n_l$ or $S$ is  small. Having passed the test there,  
it seems immune to further jeopardy. Finally,  if $n_l S$ is  
half-integer, an odd number of lines will emanate from the origin,  
one of which will go straight up to $J'=\infty$ (corresponding to nonstaggered  
odd-$n_l$ half-integer spin chains ladders, known to be gapless).

  What does the sigma model have to say about holes? 
It was shown by Shankar\cite{shankholes}, that  in the  large $S$ limit, 
  the holes in the single
 chain may be represented by spinless fermions that couple to the
 sigma model field via a gauge interaction.  
 At finite doping,  the 
fermions  render the $\theta$ term ineffective,
 wiping out the sharp distinction  between integer and half-integer chains.
 The extension of this  calculation  to ladders will tell us if 
  all ladders  have exponential decay upon doping.  

To summarize, we have considered the phase diagram of ladders with
 staggered couplings by   mapping the ladders to a sigma model
 and setting its topological coefficient 
 $\theta$ to an odd multiple of $\pi$ 
(when the model is known to be massless).  There were a few surprises:  we have examples 
here wherein coupling gapped chains 
leads to gapless chains. 
This is because there is an interplay 
between staggering and interchain coupling 
which separately destroy gaplessness, 
but together can conspire to keep  the system gapless.  
Thus  two  spin-${1\over 2}$ chains with small staggering and
 small $J'$ can remain massless.  At all these phase transitions 
 the gap will vanish  as $t^{2/3}$, (up to logarithms). 
We expect gapped states to  exhibit  linear confinement of a pair of defects
 and possibly pairing,  if the doping is macroscopic.

 It will be worth confirming these 
 predictions by  
 Monte Carlo, Density Matrix Renormalization Group, series expansions 
 and so on. The sigma model complements these approaches:   it does not do so well numerically, 
but manages to give at one stroke the phase diagram 
for any choice of $S$ and $n_l$.  
For instance, we know that on $\gamma =\pm 1$, where we have a honeycomb
ladder, each transition of a single chain 
gets tranformed into $n_l$ transitions as $J'$ is varied. 
This accumulation of critical points 
(for any spin, half-integer or otherwise) 
facilitates  extrapolation  to the ordered  state in $d=2$, although we cannot raise $n_l$ too much.

We leave it to the  
ingenuity of the experimentalists \cite{azuma} 
 to  find ladders wherein 
bonds alternate in both directions (\ref{5}), and either 
 $\gamma$ or $J'/J$, or both,  can be varied at least slightly. 
There is also the option of studying the honeycomb ladder, an extreme case of 
bond alternation ($\gamma = \pm 1$). 
Once any such a ladder is found, it will have many transitions, 
whatever be the spin. For example a honeycomb ladder with four
 legs and spin 1 will have four transitions 
as $J'$ varied, say by applying pressure.

Given the many formal similarities between the $\theta$ parameter
of the Quantum Hall Effect and  the one we have here, one can expect strong 
parallels between coupled chains and coupled Hall planes.  

Note added in proof: The phase diagrams have a nice extesnion 
to  $J'<0$, for the case where 
$J_a(n) = J(1 + \gamma (-1)^n)$, i.e., 
the staggering is only along the leg  but not along the rung direction:  if we lower 
$J'$ from 0 to $-\infty$    each transition point 
of the decoupled spin-S chain  splits into $n_l$ lines  
and all $2n_l S$ of them flow down to $J=-\infty$ and terminate 
at the gammas corresponding to the $n_l S$ transitions of the spin 2S chain.

R.S. thanks I. Affleck, E.Dagotto, M.P.A. Fisher,  S. Sachdev and D. Scalapino 
 for  useful  
conversations,   the Instituto de 
Matem{\'a}ticas y F{\'\i}sica Fundamental (C.S.I.C) for its  
hospitality when this work was commenced 
and the Institute of Theoretical Physics at Santa Barbara, 
where this work was written up. 

Work partly supported by CICYT under
contract AEN93-0776 (M.A.M.-D.), 
NSF  grants DMR 9120525,  PHY94-07194 (R.S.),
and by the 
Spanish Fund PB92-1092 
and European Community Grant ERBCHRXCT920069 (G.S.).

%
%
\begin{figure}
\caption{
A typical staggered ladder with  staggered couplings
$J (1 \pm \gamma)$ along the
horizontal legs and  $J^\prime$  along the vertical rungs.
\label{f1}
}
\end{figure}
%
%
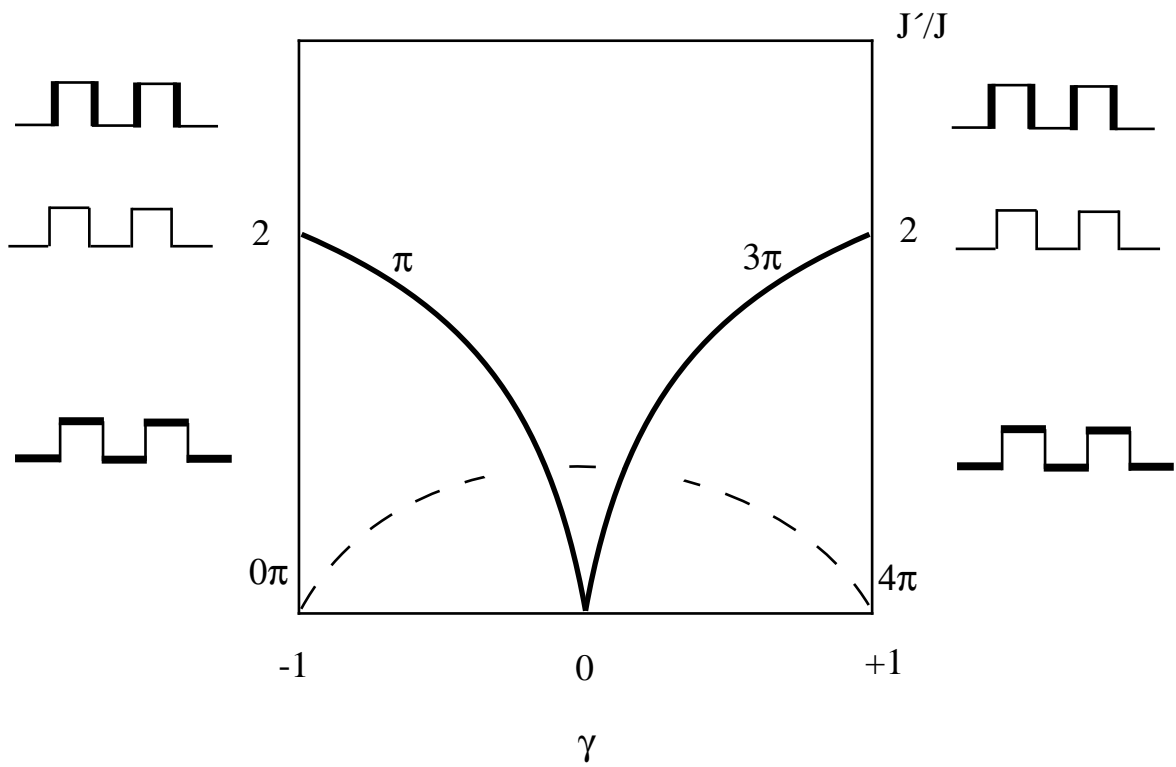
\begin{figure}
\caption{
Phase diagram for staggered ladders with $n_l = 2$ and spin ${1\over 2}$.
The solid lines represent critical lines corresponding to the $\theta$-parameter being
an odd multiple of $\pi$, (the dashed line is just indicative for counting the number of
critical lines between $\gamma =-1$ and $\gamma =+1$.)
On the margins we show the snake patterns associated to criticality (non-staggered)
and to gaped phases (staggered).
\label{f2}
}
\end{figure}
%
%
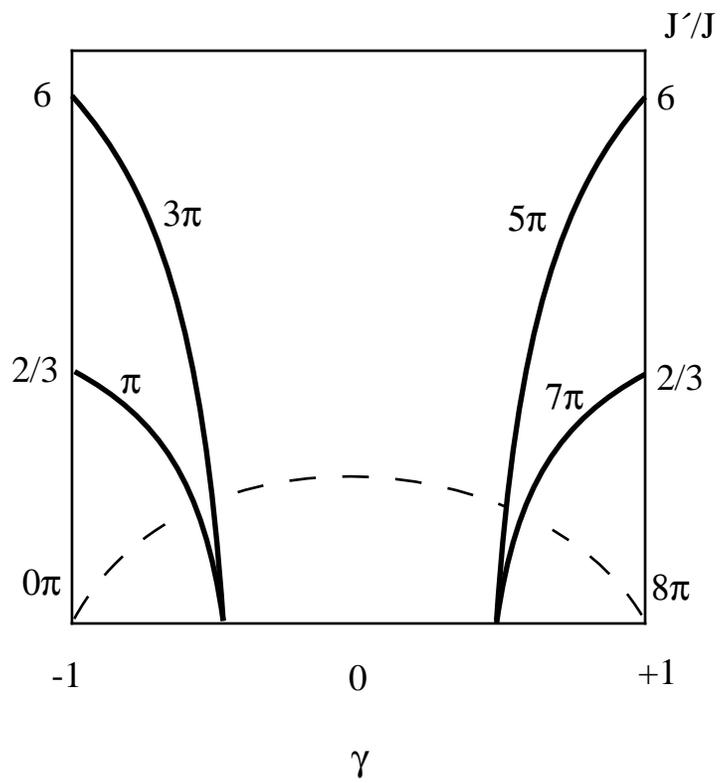
\begin{figure}
\caption{
Phase diagram for staggered ladders with $n_l = 2$ and spin $1$.
See Fig. 2  for similar explanations.
\label{f3}
}
\end{figure}

\end{document}